\documentclass[a4paper]{aa}

\usepackage{graphicx}
\usepackage{amsfonts}
\usepackage{amsmath}

\graphicspath{{/home/moraux/Nbody2/Vbd=kVstar/}
{/home/estelle/NBody/NBody2/Vbd=kVstar/}}

\newcommand{\msun}{M_{\odot}}
\newcommand{\bd}{brown dwarf }
\newcommand{\bds}{brown dwarfs }
\newcommand{\pl}{Pleiades }

\begin{document}

\title{Kinematics of stars and brown dwarfs at birth}

\institute{%
Institute of Astronomy, Madingley Road, Cambridge, CB3 0HA, UK
}

\author{E. Moraux \and C. Clarke}

\offprints{moraux@ast.cam.ac.uk}

\date{}

\authorrunning{E. Moraux \& C. Clarke}
\titlerunning{Kinematics of stars and brown dwarfs at birth}

\abstract{

We use numerical N-body simulations in order to test whether the
kinematics of stars and brown dwarfs at birth depend on mass. In
particular we examine how initial variations in velocity dispersion
can affect the spatial distribution of stellar and substellar objects
in clusters. We use 'toy' N-body models of a Pleiades-like cluster in
which brown dwarfs have their own velocity dispersion
$\sigma_{V_{\text BD}}$ which is $k$ times larger than the stellar
one.

We find that in order to match the broad agreement between the brown
dwarf fraction in the field and in the Pleiades, the velocity
dispersion of brown dwarfs at birth has to be less than twice the
stellar velocity dispersion, i.e. cannot exceed a few km/s in the
Pleiades cluster. In order to discern more subtle differences between
the kinematics of \bds and stars at birth, our simulations show that
we need to look at clusters that are much less dynamically evolved
than the Pleiades. One might especially seek evidence of high velocity
brown dwarfs at birth by examining spatial distribution of stars and
brown darfs in clusters that are about a crossing timescale old.

}

\maketitle


\section{Introduction}

So far, most studies of the star formation process have dealt with the
formation of stellar systems but with the recent discovery of brown
dwarfs in 1995 (Nakajima et al. 1995; Rebolo et al. 1995) new
perspectives have opened regarding the formation of condensed objects
in molecular clouds. Today more than a hundred brown dwarfs (BDs) in
various environments are known. However, their mode of formation is
still controversial and the theoretical framework describing the
stellar and substellar formation process(es) is not completely
satisfactory. Regions of high density ($n(H_{2})\sim 10^{7}$
cm$^{-3}$) in molecular clouds are needed to form proto-\bds but then
for their mass to remain substellar their reservoir of gas has to be
small or the accretion not very efficient.

Two main competing scenarios have been proposed so far to account for
the formation of substellar objects. One assumes that brown dwarfs
form like solar-mass stars by gravitational collapse of small, dense
molecular cloud core and subsequent accretion. The supporting argument
is that in the opacity limited regime the Jeans mass can be as low as
a few Jupiter masses (Low \& Lynden-Bell 1976). The alternative view
assumes that brown dwarfs are ejected ``stellar embryos'' as proposed
by Reipurth \& Clarke (2001). In this scenario, molecular cloud cores
fragment to form unstable protostellar multiple systems which decay
dynamically. The lowest mass fragments are ejected from their birth
place and deprived of surrounding gas to accrete remain substellar
objects.

The brown dwarf properties predicted by these two different formation
scenarios may in principle be quite different. In the former case,
both stars and brown dwarfs form predominantly as single or binary
($N=2$) systems; in this case there are no obvious reasons why
properties such as the binary fraction or kinematics should depend on
mass.  In the latter scenario, by contrast, the dominant formation
mechanism (again for both stars and brown dwarfs) is in small $N$
($>2$) clusters, and the gravitational interplay that precedes the
break up of the system into stable entities implies a potentially
strong mass dependence for resulting properties like the binary
fraction and kinematics. In particular it has been suggested (Reipurth
and Clarke 2001) that low mass objects (e.g. brown dwarfs) ejected
from such clusters would have a higher velocity dispersion than higher
mass objects.

Reipurth and Clarke's initial suggestion - that brown dwarfs in star
forming regions may have a detectably higher velocity distribution
from stars - has {\it not} been borne out by radial velocity studies
(Joergens \& Guenther 2001). Meanwhile, successive simulations have
modified the predictions of small $N$ clusters models. Delgado et
al. (2003) and Sterzik and Durisen (2003) have emphasised that, in
their simulations, the main difference in velocity dispersion is
between single stars and binaries, and that brown dwarfs attain rather
larger velocities - with respect to their parent cores - because they
are more likely to be ejected as single objects. This dependence of
ejection speed on binarity may readily be understood, since one binary
is typically formed in each cluster in these simulations: this binary
is able to eject the remaining stars from the cluster by sling-shot
gravitational encounters, whilst itself remaining close to the center
of mass of the natal cluster. In the turbulent fragmentation
calculations of Bate, Bonnell and Bromm (2003) and Delgado, Clarke and
Bate (2004), by contrast, more than one binary is formed per cluster
and so binaries are able to eject each other from the natal
cluster. Consequently, in these simulations, the kinematics of the
resulting objects do not depend strongly on either mass {\it or}
binarity.

Evidently, the relative kinematics of stars and brown dwarfs and of
single stars and binaries can shed some light on the conditions in
star forming cores and could ultimately answer the question of whether
stars (and brown dwarfs) are formed as isolated single and binary
systems, as small $N$ aggregates containing typically one binary or as
aggregates containing more than one binary. [Note that this question
is not easy to answer by direct observations, since the timescale for
the break up of putative small clusters implies that this process
occurs in the deeply embedded phase. However, high resolution imaging
of the driving sources of Herbig Haro objects by Reipurth (2000)
suggests that the multiplicity of stars in deeply embedded regions is
indeed high].

Direct observations of the kinematics of young stars and brown dwarfs
is unlikely to be fruitful however. The differences in velocity
dispersion predicted by theoretical models are small (of the order of
a km/s). When one bears in mind that these velocities are measured
with respect to star forming cores, which are in themselves in
relative motion at $\sim1$ km/s, it is unsurprising that the study of
Joergens and Guenther (2001) - involving small numbers of objects,
with velocity resolution of $\sim0.2$ km/s and rather small dynamic
range in mass - did not detect any differences.

In this paper, we propose another approach that could potentially
detect any mass dependence of the kinematics of stars (and brown
dwarfs) at birth. Here we examine the statistical consequences of such
an effect on the spatial distribution of stars and brown dwarfs in
clusters. This approach has the advantage that one can work with large
samples of stars and brown dwarfs, whose positions and masses are
known with high accuracy. On the other hand, we cannot predict how
initial variations in velocity dispersion affect the spatial
distributions of stars and brown dwarfs in a cluster at a given age
without further, N-body, modeling.  This is partly because two body
relaxation leads to mass segregation in older clusters, even in the
absence of a mass dependent initial velocity dispersion.  Our purpose
in this paper, therefore, is to use `toy' N-body models (in which brown
dwarfs are introduced with a velocity dispersion that is a variable
multiple of the stellar velocity dispersion) in order to establish
under what circumstances could one detect a higher velocity dispersion
at birth for low mass objects. We stress that this toy models for the
kinematics is not supposed to correspond to the outcome of any
particular numerical star formation model but is designed to provide a
ready parameterization of the problem. We also underline that in no
models are any sudden discontinuities in kinematic properties expected
at the hydrogen burning mass limit.

We use the Pleiades as the testbed for our calculations. This is
because the brown dwarf population of the Pleiades has been the
subject of intensive scrutiny in recent years (Moraux et al. 2003,
Dobbie et al. 2002, Pinfield et al. 2000, Zapatero-Osorio et al. 1999,
Bouvier et al. 1998) so that the present day mass function in this
cluster is reasonably well constrained. We shall proceed by first
placing an upper limit on the initial velocity dispersion of brown
dwarfs in the Pleiades, based on the broad similarity between the
normalization of stars to brown dwarfs in the Pleiades and in the
field, which limits the number of brown dwarfs that can have left the
cluster to date. We shall then explore whether the radial distribution
of brown dwarfs in the cluster can place meaningful limits on their
initial velocity distribution.

\section{Numerical simulations}

We performed numerical simulations of the dynamical evolution of a
Pleiades-like cluster using the code {\sc Nbody2} (Aarseth 2001) on a
Sun workstation. This code is an algorithm for direct integration of
N-body problem based on the neighbour scheme of Ahmad \& Cohen (1973)
and employs a softened potential $\phi$ of the form
\begin{equation}
  \phi = -\frac{m}{(r^{2}+\epsilon^{2})^{1/2}}
\end{equation}
to reduce the effects of close encounters.

The model of cluster we used is defined as follows. At time $t=0$ the
stellar density $n(r,t)$ conforms to Plummer's model
\begin{equation}
  n(r,0) = \frac{3}{4\pi r_{0}^{3}} N  \left[ 1 + \left(
    \frac{r}{r_{0}} \right)^{2} \right]^{-5/2}
\end{equation}
where $N=1900$ is the number of cluster members and $r_{0}=2.2$
pc is a scale factor determining the dimensions of the cluster. It is
related to the half-mass radius $r_{h}$ by $r_{h}\simeq1.3\,
r_{0}=2.86$ pc (Aarseth \& Fall 1980). This leads to an overall
initial central density $n(0,0)=42.6$ objects/pc$^{3}$. Initially,
the stellar population is assumed to be in virial equilibrium with a
velocity distribution everywhere isotropic (cf. Aarseth, H\'enon \&
Wielen 1974 for a practical scheme for the generation of the initial
positions and velocities). Note that the true initial conditions
of a cluster are not very well known and are likely to be very
complex. Isothermal models are sometimes preferred to describe open
cluster density initial states, however Plummer models are also used
and have already been proved to reproduce reasonably well the Pleiades
cluster (Kroupa et al. 2001). In section~\ref{radii} we compare our
results to observational data and find a reasonable agreement.

The system is assumed to be isolated. No external potential is
included but any object which reaches the cluster tidal radius would
in reality be stripped off by the galactic tide and lost by the
cluster.

For simplicity and in order to focus on how the initial kinematics
affects the the spatial distribution of the cluster population, our
model does not include gas. We assume that the gas has already left
the cluster when we start the simulations and thus the original
cluster may have expanded and lost a large fraction of its primordial
members because of the change of the gravitational potential (Adams
2000, Boily \& Kroupa 2003a, 2003b). This explains in particular why
our initial system is not as concentrated as e.g. the cluster models
used by Kroupa et al. (2001) to reproduce the Pleaides. We shall
discuss how the presence of gas would affect our results later on.

The smoothing length employed for the gravitational interactions on
small scales is $\epsilon=5\times10^{-4}\, r_{0}\sim200$ A.U.. To
justify the choice of this value, one can estimate the rate of
encounters per star closer than $\epsilon$ by
\begin{equation}
  f = 4\sqrt{\pi}n \left( \sigma_{V}\epsilon^{2} +
\frac{Gm\epsilon}{\sigma_{V}} \right)
\end{equation}
(e.g. Binney \& Tremaine 1987) where $n$ is the local stellar density,
$\sigma_{V}$ is the velocity dispersion and $m$ is the stellar mass,
assumed to be the same for all stars. Assuming $n$ does not change
with time, we find the probability for a star located at $r=0$ to have
encounters closer than $\epsilon$ is $\sim9$\% in 120 Myr -- wich is
about the age of the \pl (Stauffer et al. 1998, Mart\'in et
al. 1998). Since the stellar density decreases with time, this value
is an upper limit indicating $\epsilon\sim200$ A.U. is appropriate for
our simulations.

The mass of the $N=1900$ objects are distributed over a three
power-law mass function
\begin{equation}
  \xi(m) = \frac{dn}{dm} \propto m^{-\alpha_{p}}\,,\,\, p\in[1..3]
\label{imf}
\end{equation}
with
\begin{eqnarray*}
  \alpha_{1} &=& 0.6 \,\,{\textrm{ for }} m_{1,inf}=0.01\le m
  \le m_{1,sup}=0.3 \msun, \\ 
  \alpha_{2} &=& 1.3 \,\,{\textrm{ for }} m_{2,inf}=0.3\le m
  \le m_{2,sup}=1.0 \msun, \\ 
  \alpha_{3} &=& 2.3 \,\,{\textrm{ for }} m_{3,inf}=1.0\,\le m 
  \le m_{3,sup}=10.0 \msun,
\end{eqnarray*}
which corresponds to the \pl mass function determined by Moraux et
al. (2003). This MF estimate has not been corrected for binarity which
means that this mass function corresponds to the {\it system} mass
function. Likewise, our simulations do not include any treatment of
binarity. Using a softened potential inhibits binary formation and
each object issued from this mass distribution is considered as a
single object. This choice can be justified by results obtained
by de la Fuente Marcos \& de la Fuente Marcos (2000) who performed
simulations to investigate the dynamical evolution of substellar
population in cluster with Aarseth's {\sc Nbody5} code (Aarseth 1985).
Some of their models include a population of hard binaries, which are
the most important binary stars in term of dynamics, but the results
are all very similar (see their fig.2). This suggests that close two
body encounters are not too important and that a softened potential
$\phi$ with an adequate $\epsilon$ can be used.

In practice, the mass of each member $i$ is readily obtained by
\begin{equation}
  m_{i} = m_{p,sup}^{-(\alpha_{p}-1)} - (i-1) g_{N_{p}}
\end{equation}
with
\begin{equation*}
  g_{N_{p}} = (m_{p,sup}^{-(\alpha_{p}-1)} -
  m_{p,inf}^{-(\alpha_{p}-1)})) / (N_{p} - 1)
\end{equation*}
where $N_{p}$ is the number of objects having a mass between
$m_{p,inf}$ and $m_{p,sup}$. A convenient representation for this
distribution is the mass-generating function
\begin{equation}
  m(X) = \left\{
	\begin{array}{ll}
	\frac{0.3}{[1-1.35(X-0.55)]^{-2.5}} & \textrm{if } 0<X\le 0.55,\\
	\frac{1}{[1-1.45(X-0.85)]^{3.33}} & \textrm{if } 0.55<X\le 0.85,\\
	\frac{0.24}{[1-0.99X]^{0.77}} & \textrm{if } 0.85<X\le 1,\\
	\end{array}
  \right.
\end{equation}
where $X$ is distributed uniformly in [0,1]. The intervals here are
the same as the mass intervals in Eq.~(\ref{imf}). This function has been
computed in the way described in Kroupa et al. (1991).

In such a model, \bds ($0.01\le m\le 0.08\msun$) constitute 25\% by
number but only less than 3\% by mass. Initially these proportions are
the same everywhere in the cluster, i.e. there is no mass segregation,
so that stellar and substellar objects have the same radial
distributions. The initial density of stars and brown dwarfs in
the center are $n_{\star}(0,0)=32$ and $n_{\rm{BD}}=10.6$
objects/pc$^{3}$ respectively.

In our toy model we assume that substellar objects represent a
peculiar population having its own velocity dispersion
$\sigma_{V_{\rm{BD}}}$. We arbitrarily choose
\begin{equation}
  \sigma_{V_{\rm{BD}}}=k\times\sigma_{V_{\star}}
\end{equation}
initially where $k\in [1.0-3.0]$ and $\sigma_{V_{\star}}$ is the
stellar velocity dispersion.

Then we let the cluster evolve dynamically under the effect of
gravitational interactions between members over 12 crossing
times. Since $t_{cr}\simeq 10$ Myr for the Pleiades (Pinfield et
al. 1998), this corresponds to about the age of the cluster, i.e. 120
Myr. Thus, we can follow the evolution of the star and \bd population
within the cluster from their birth to the age of the Pleiades
depending on their initial kinematics.

Note that we only present the results for one set of initial
conditions but we also performed other simulations using different
numbers of objects $N$, which means in particular using various seed
numbers to initialize the cluster model. The results are similar to
those described in the following sections.

\section{Results and discussion}
\label{results}

\subsection{Characteristic radii}
\label{radii}

Over the timescale of the simulations the half mass radius does
not change much through the simulations. It goes up to $\sim3.2-3.4$
pc after 12 crossing times which is consistent with observational
results obtained for the Pleiades ($r_{h}=3.6$ pc; Pinfield et
al. 1998). This result does not depend on the initial substellar
velocity dispersion which is indeed expected considering the fact that
\bd represent only 3\% of the cluster mass.

A nominal core radius $r_{c}$ is calculated in {\sc Nbody2} using the
definition of the density radius given by Casertano \& Hut (1985)
slightly modified in order to obtain a convergent result using a
smaller sample ($n\simeq N/2$). It is determined by the rms expression
(Aarseth 2001)
\begin{equation}
  r_{c}= \left( \frac{\sum_{i=1}^{n} |\mathbf{r}_{i} -
\mathbf{r}_{d}|^{2} \, \rho_{i}^{2}}{\sum \rho_{i}^{2}} \right)^{1/2},
\end{equation}
where $\mathbf{r}_{i}$ is the three-dimensional position vector of the
$i$th star and $\mathbf{r}_{d}$ denotes the coordinates of the density
centre. The density estimator $\rho_{i}=3\,M_{5}/(4\pi r_{6}^{3})$ is
defined with respect to the sixth nearest particle $r_{6}$ and takes
into account the total mass ($M_{5}$) of the five nearest
neighbours. [See Casertano \& Hut (1985) for definitions of density
centre and density estimator.] In our simulations, $r_{c}=1.3$ pc at
$t=0$ and $r_{c}\sim0.8$ pc at $t=12\,t_{cr}\simeq 120$ Myr. However,
to be compared to the observational core radius $R_{c}$ these values
have to be divided by $\sim0.8$ (Heggie \& Aarseth 1992) and we obtain
$R_{c}$ between 1.6 and 1 pc. For the Pleiades, Raboud \& Mermilliod
(1998) found $R_{c}=0.6\degr=1.3$ pc for a cluster distance of 125 pc
which is reasonably consistent with our model.

\begin{figure}[htbp]
  \leavevmode 
  \includegraphics[width=0.45\textwidth]{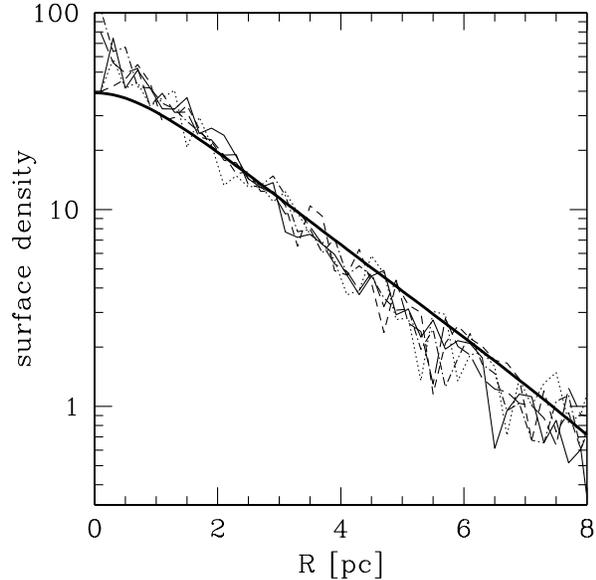}
  \caption{The projected radial density profiles for $0.1\le
    m\le3\msun$. The different thin curves correspond to different
    values of $\sigma_{V_{\rm{BD}}}=k\times\sigma_{V_{\star}}$ with
    $k\in [1.0-3.0]$, the symbols are the same than in
    Fig.~\ref{cumul120}. The thick curve correspond to the
    best-fitting King profile of observational data for the Pleiades
    from Pinfield et al. (1998, their fig.~2.)}
  \label{density}
\end{figure}

We also compare the projected radial density profiles obtained at
$t\simeq 120$ Myr to the observed Pleiades profiles for $0.1\le m \le
0.3\msun$ (see Fig.~\ref{density}) and we find a reasonable agreement.

\subsection{Dynamical evolution of the substellar population}

\begin{figure}[htbp]
  \leavevmode
  \includegraphics[width=0.45\textwidth]{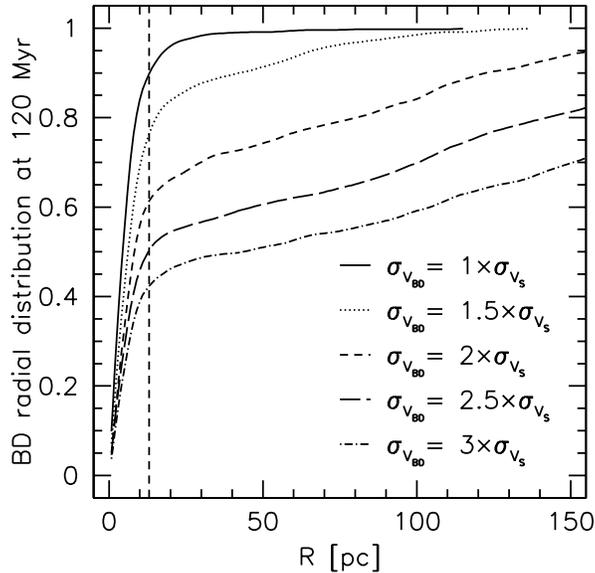}
  \caption{Cumulative radial distribution of \bds after $\sim 120$ Myr
    calculated by our {\sc Nbody2} numerical simulations of the
    dynamical evolution of a Pleiades-like cluster. The different
    curves correspond to the different values of the substellar
    initial velocity dispersion, $\sigma_{V_{\rm{BD}}} = k \times
    \sigma_{V_{\star}}$ with $k\in [1.0-3.0]$ ($k$ increases from top
    to bottom). The vertical dashed line represents the extend of the
    Pleiades cluster.}
  \label{cumul120}
\end{figure}

Figure~\ref{cumul120} illustrates the effect of the initial velocity
dispersion of \bds on their spatial distribution at the age of the \pl
cluster ($\sim 120$ Myr). The vertical dashed line corresponds to the
tidal radius $r_{t}$ of the \pl ($\sim 13$ pc, Pinfield et al. 1998).
All the objects located at larger radii are in reality lost by
the cluster.

If the stellar and substellar initial velocity dispersion are similar
($k=1.0$), then as many \bds as low mass stars have been lost after
120 Myr (about 10\%). This result is perfectly consistent with the
simulations performed by de la Fuente Marcos \& de la Fuente Marcos
(2000), which indicates in particular that our choice of the
softening parameter $\epsilon$ is adequate. Like the stars, these
brown dwarfs have diffused as a result of successive two body
interactions to beyond the notional tidal radius. However, as soon as
$k\ge2.0$ then 40\% or more of the initial \bd population have left
the cluster.

\begin{figure}[htbp]
  \leavevmode
  \includegraphics[width=0.45\textwidth]{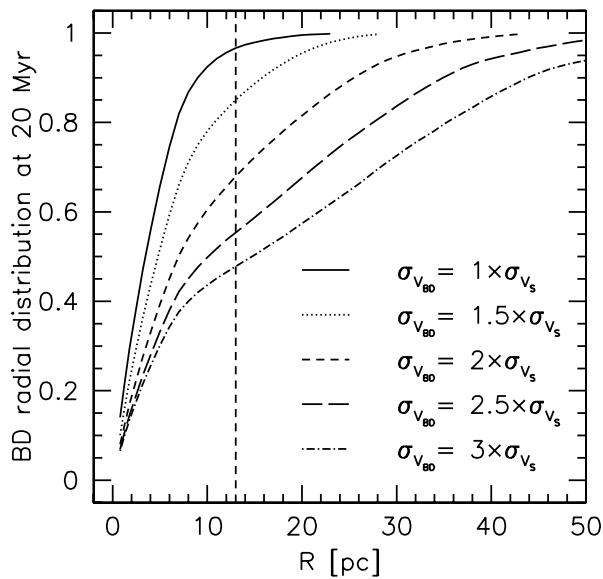}
  \caption{Cumulative radial distribution of \bds after $\sim$ 20 Myr
    for different values of initial substellar velocity dispersions
    $\sigma_{V_{\rm{BD}}}=k\times\sigma_{V_{\star}}, k\in [1.0-3.0]$.}
  \label{cumul20}
\end{figure}

If it does happen, this loss of \bds occurs quite quickly, in a
couple of crossing times, i.e. a few tens of Myr. We note from
Fig.~\ref{cumul20} that the depletion rates obtained after only 20 Myr
are already important and of the same order as those obtained after
120 Myr. This means in particular that our results do not depend
much on the age of the cluster and are not affected by the errors in
open cluster age determinations, typically in the range of 30 to
50\%. This can be explained by the fact that if an object has a
velocity larger than the cluster escape velocity $v_{esc}$ it will be
lost after a few crossing times $t_{cr}$. The depletion rates found at
20 Myr correspond indeed to about the fraction of \bds initially
unbound in our Plummer distribution. For our Pleiades-like model, we
have $t_{cr}\simeq10$ Myr and $v_{esc}=\sqrt{2}v_{vir}=1.6$ km/s, with
the virialised velocity $v_{vir}=1.15$ km/s. Thus, if \bds form
following the ejection scenario proposed by Reipurth \& Clarke (2001)
with a resulting average velocity larger than the cluster escape
velocity or with a velocity dispersion larger than a few km/s, then a
large number of substellar objects will be lost relatively quickly.

The fact that basically all the initially unbound objects are lost by
the cluster is quite straightforward and constitutes a robust result
that does not depend much on the initial conditions, like the IMF or
the initial spatial distribution. Evidently the actual fractions of
stars and brown dwarfs that would be bound at the start of our
simulations would partly depend on the prior dynamical history of the
cluster when it contained significant quantities of gas, an issue that
is beyond the scope of the present paper. We note however that changes
in the cluster potential due to gas loss would not affect the
kinematic properties of the stars and brown dwarfs {\it
differentially} and in this study we focus on the possible observable
consequences of stars and brown dwarfs having different kinematical
properties at birth. 

However, an important observational constraint is that the \pl mass
function is found to be similar to those of several star forming
regions (where the gas is still present), such as the Trapezium
(Muench et al. 2002) or IC~348 (e.g. Luhman et al. 2003), and of the
field (cf. Moraux et al. 2003). This indicates that it is still
representative of its initial mass function and that the dynamical
effects did not affect the shape of the IMF in 120 Myr. This means in
particular that there was {\it no} significant preferential escape of
\bds in the Pleiades. Therefore, the substellar initial velocity
dispersion must be less than twice the stellar velocity dispersion and
cannot exceed a few km/s.

\subsection{Radial distribution}

As discussed above, the vast majority of objects has a typical
velocity of a few km/s and remains bound to Pleiades-like clusters
after $\sim100$ Myr. However, if \bds still have a larger dispersion
velocity than that of stars, even if only slighty larger, we may hope
to find a signature of their ejection by looking at their radial
distribution.

\begin{figure}[htbp]
  \leavevmode
  \includegraphics[width=0.45\textwidth]{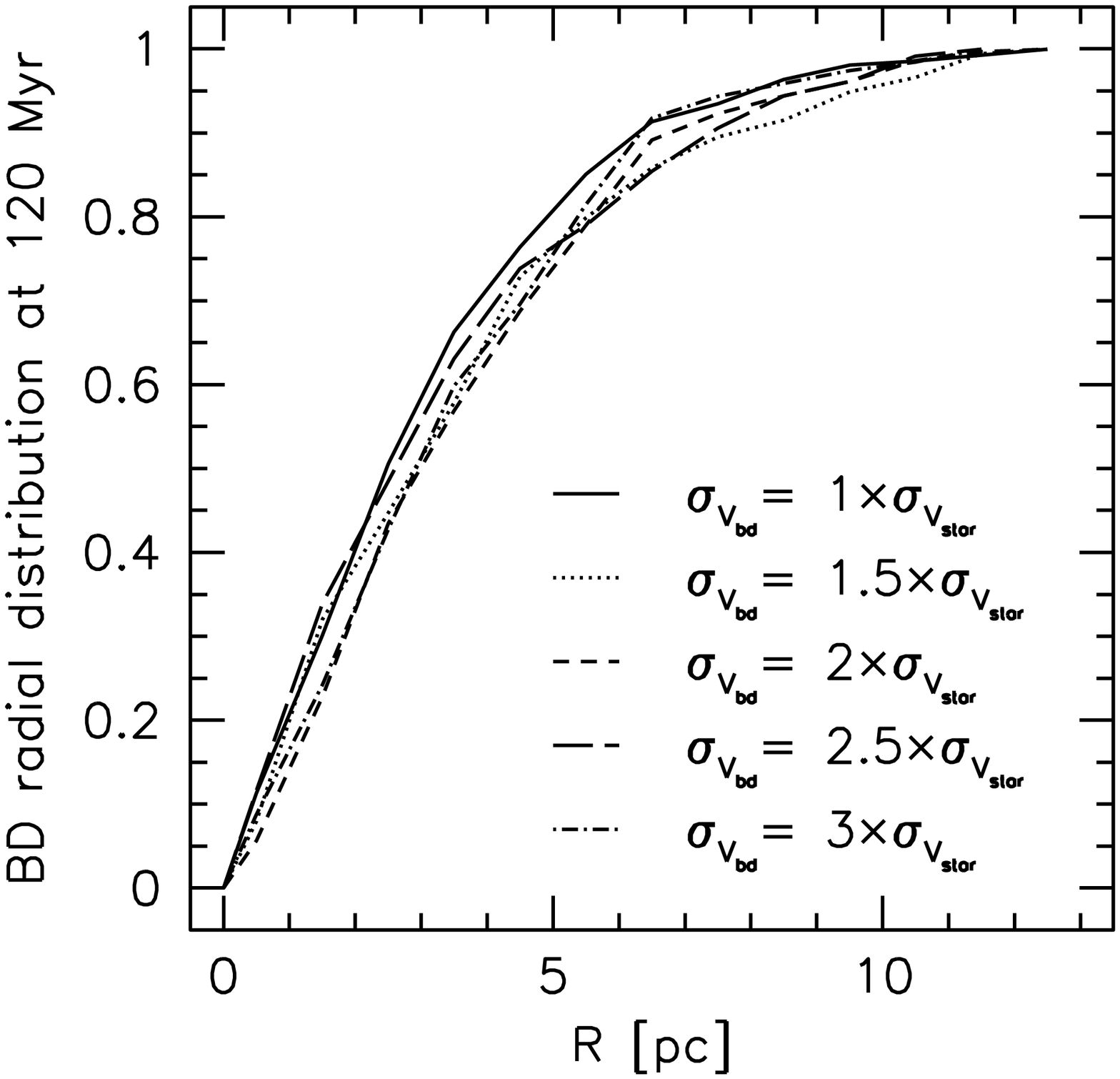}
  \caption{Cumulative radial distribution of \bds inside the cluster
    after $12 t_{cr}\sim 120$ Myr for several initial substellar
    velocity dispersions
    $\sigma_{V_{\rm{BD}}}=k\times\sigma_{V_{\star}}, k\in [1.0-3.0]$.}
  \label{cumul_cut120}
\end{figure}

Figure~\ref{cumul_cut120} shows the effect of the initial velocity
dispersion on the \bd radial distribution for a Pleiades-like cluster
after $12 t_{cr} \sim 120$ Myr. The number of objects inside a sphere
having a radius $R=13$ pc has been normalised to one. The radius has
been chosen to correspond to the Pleiades tidal radius so that the
shown radial distributions represent what an observer would obtain if
he studied the substellar population in the cluster.

Strikingly, we find that at the age of the Pleiades, the spatial
distribution of brown dwarfs in the cluster provides {\it no}
information about the velocity dispersion of brown dwarfs at
birth. This is because those brown dwarfs whose velocity exceeds the
escape velocity of the cluster will have already left the cluster (on
a crossing timescale $\sim 10$ Myr) and those that remain have a
velocity distribution (and hence spatial distribution) that is very
similar to that of the low mass stars.

\begin{figure}[htbp]
  \leavevmode
  \includegraphics[width=0.45\textwidth]{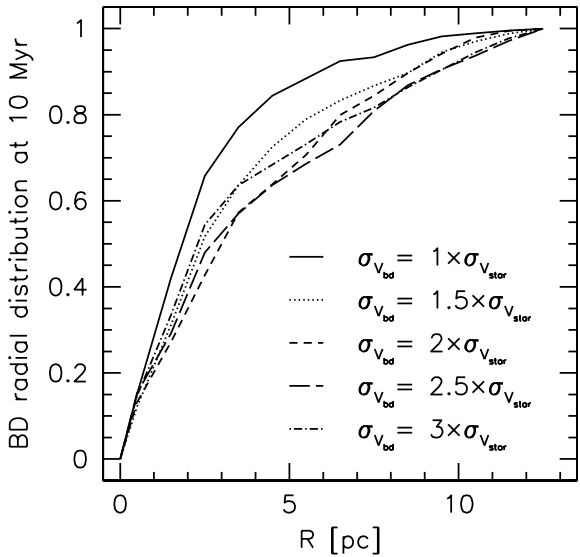}
  \caption{Cumulative radial distribution of \bds inside the cluster
    at $t=1 t_{cr}\sim 10$ Myr for several initial velocity
    dispersions $\sigma_{V_{\rm{BD}}}=k\times\sigma_{V_{\star}}, k \in
    [1.0-3.0]$.}
  \label{cumul_cut10}
\end{figure}

\begin{figure}[htbp]
  \leavevmode
  \includegraphics[width=.45\textwidth]{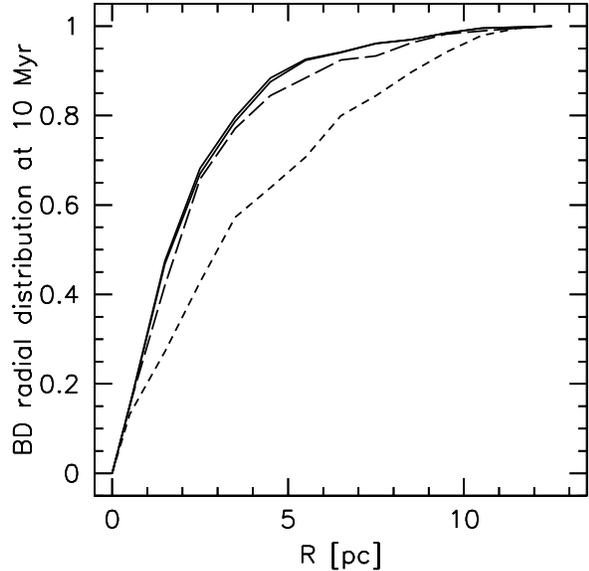}
  \caption{Radial distribution of low mass stars ($0.08-0.5\msun$;
    solid line) and brown dwarfs (dashed lines) inside the cluster
    after $1t_{cr}\sim 10$ Myr. The long-dashed line corresponds to
    $\sigma_{V_{\rm{BD}}}=\sigma_{V_{\star}}$ and the short-dashed
    line to $\sigma_{V_{\rm{BD}}}=2\times\sigma_{V_{\star}}$.}
  \label{bd_star10}
\end{figure}

In order to find possible evidence of a population of brown dwarfs
with high velocities at birth, it is instead necessary to examine
clusters that are only about a crossing time old. This expectation is
borne out by Figures~\ref{cumul_cut10} and~\ref{bd_star10} which
compare the normalised distributions of stars and brown dwarfs within
a Pleiades-like cluster at an age of $10$ Myr for various ratios of
$\sigma_{V_{\rm{BD}}}$ to $\sigma_{V_{\star}}$.  Evidently, at this
age, two-body relaxation is ineffective in producing a more diffuse
brown dwarf distribution (as evidenced by the fact that the brown
dwarf and stellar distributions are very similar at this age if
$\sigma_{V_{\rm{BD}}} = \sigma_{V_{\star}}$). Thus any differences in
the spatial distribution of stars and brown dwarfs at such a young age
would be indicative of different velocity distributions at birth.

Figure~\ref{cumul_cut10} shows the effect of the initial velocity
dispersion on the \bd radial distribution for a Pleiades-like cluster
after $1 t_{cr} \sim 10$ Myr. We note that as soon as the \bd
velocity dispersion is larger than that of stars the substellar radial
distributions are different from the one obtained for
$\sigma_{V_{\rm{BD}}}=\sigma_{V_{\star}}$.

Figure~\ref{bd_star10} compares the spatial distribution of low mass
stars to that of brown dwarfs for two different values of
$\sigma_{V_{\rm{BD}}}$ after 10 Myr. When the stellar and substellar
velocity dispersions are similar then the radial distributions are
also the same, whereas this is not the case for
$\sigma_{V_{\rm{BD}}}=2\times \sigma_{V_{\star}}$. In that case about
90\% of the low mass stars ($0.08\le m\le 0.5\msun$) are located at
less than 5 pc from the cluster center compared with only 65\% of the
\bds at the same location. We have to reach a radius of $\sim 9$ pc to
include 90\% of the substellar population.

In the presence of gas, the results would be similar. One would
expect gas expulsion due to e.g. photoionisation or stellar winds to
occur within a few Myr, i.e. on a timescale less than the crossing
timescale of the cluster, and for the stars and brown dwarfs to
respond to such gas loss on about a crossing timescale. If brown
dwarfs have a larger velocity dispersion than stars, we therefore
expect to find a signature of this initial kinematics in studying the
spatial distribution of stars and brown dwarfs in clusters that are
about a crossing timescale old.

\section{Conclusion}

We have shown that the observed similarity between the brown dwarf to
star ratio in the Pleiades, in star forming regions and in the field
implies that the velocity dispersion of brown dwarfs at birth cannot
exceed $2\times$ the stellar velocity dispersion in the Pleiades. This
imposes an upper limit on the brown dwarf velocity dispersion at birth
of a few km/s. Thus either brown dwarfs are not dynamically ejected
from their natal cores or else the ejection velocities with respect to
their natal cores is low. Such a velocity dispersion limit would rule
out the common incidence of very hard encounters (i.e. at $<$ 4 A.U.,
Delgado et al. 2004) in the natal cores, but would be consistent with
the results of current hydrodynamical simulations (Delgado et
al. 2003, Bate et al. 2003). We stress that the velocities quoted here
are limits on the 3D velocity dispersion for a Maxwellian distribution
and so do not necessarily rule out a tail of objects with
significantly higher velocities (such, for example, as the (not
substellar) object PV Ceph, for which a velocity of $20$ km/s has
recently been claimed (Goodman \& Arce 2003).

We have shown that the radial density profile of brown dwarfs and
stars in the Pleiades provides {\it no} information about the velocity
dispersion of brown dwarfs at birth. This is because any brown dwarfs
with velocities greater than the escape velocity of the cluster would
have long ago escaped the cluster and thus the residual brown dwarf
population has a velocity distribution - and hence spatial
distribution - that is very similar to the stars.

We therefore conclude that if we seek {\it positive} evidence for a
modestly higher velocity distribution of brown dwarfs at birth, we
need to look at clusters that are significantly different from the
Pleiades. There are two possibilities here. Firstly, the upper limits
we have derived could still imply a significant preferential loss of
brown dwarfs from loosely bound regions.  For example, Preibisch et
al. (2003) suggested that this could be the cause of the small
fraction of \bds observed in IC348 for which the escape velocity is
only $\sim 0.8$ km/s (Herbig 1998). Kroupa \& Bouvier (2003) found
also that the ejection mechanism could explain a part of the deficit
of substellar objects in the Taurus association. Secondly, one might
seek evidence of high velocity brown dwarfs at birth by instead
examining clusters that are about a crossing timescale old. In this
case, the brown dwarfs would be expected to have a more extended
distribution than the stars, which could {\it not}, at such a young
age, be confused with the signature of two body relaxation. Even if
the brown dwarf velocity dispersion exceeds that of stars by only
50\%, the spatial distributions of the two populations are quite
different at this age (contrast dashed lines in Fig.~\ref{cumul_cut10}
with solid line in Fig.~\ref{bd_star10}) Thus brown dwarf surveys in
clusters of various ages and degrees of richness will afford further
opportunities to constrain the velocities of brown dwarfs at birth.


\begin{acknowledgements}

We thank S.~Aarseth for allowing us access to his N-body codes.

This work has been done under the auspices of the EC-RTN {\it ``The
formation and evolution of young open clusters''} and a large
programme at CFHT is conducted to survey PMS clusters with wide-field
cameras (CFH12K and MegaCam) to test our predictions.

\end{acknowledgements}



\end{document}